\documentclass[letter]{aa}

\usepackage[T1]{fontenc}
\usepackage{graphicx}
\usepackage{amsmath}
\usepackage{txfonts}
\usepackage{siunitx}
\usepackage{xspace}
\usepackage{booktabs}
\usepackage[hidelinks]{hyperref}

\DeclareSIUnit{\mag}{mag}
\DeclareSIUnit{\dex}{dex}
\DeclareSIUnit{\mmag}{mmag}
\DeclareSIUnit{\jansky}{Jy}
\DeclareSIUnit{\pc}{pc}
\DeclareSIUnit{\um}{\ensuremath{\muup}m}

\newcommand{\ice}{\textsc{Ice}\xspace}
\newcommand{\taumax}{\ensuremath{\tau_{3.0}^{\max}}\xspace}
\newcommand{\lam}{\ensuremath{\Lambda(W_1 - I_1)}\xspace}
\newcommand{\ak}{\ensuremath{A_{K_S}}\xspace}
\newcommand{\akz}{\ensuremath{A_{K_S}^{\mathrm{th}}}\xspace}
\newcommand{\xii}{\ensuremath{x_{12}}\xspace}
\newcommand{\gnull}{\ensuremath{G_0}\xspace}

\newcommand{\nstars}{\num{43009}}
\newcommand{\nicefree}{\num{26687}}
\newcommand{\rak}{0.052}
\newcommand{\nboot}{300}
\newcommand{\thrlow}{0.67}
\newcommand{\thrhigh}{1.15}
\newcommand{\xbinlow}{-0.19}
\newcommand{\xbinhigh}{+0.41}
\newcommand{\slopex}{0.90}
\newcommand{\slopelo}{0.72}
\newcommand{\slopehi}{1.07}

\newcommand{\slopeslow}{0.20}
\newcommand{\slopeshigh}{0.28}

\newcommand{\xrange}{1.1}
\newcommand{\bintable}{1 & $-0.19$ & $0.67^{+0.04}_{-0.03}$ & $0.28^{+0.02}_{-0.02}$ & 621 \\
2 & $-0.08$ & $0.83^{+0.02}_{-0.06}$ & $0.24^{+0.02}_{-0.03}$ & 752 \\
3 & $+0.00$ & $0.86^{+0.03}_{-0.05}$ & $0.20^{+0.02}_{-0.02}$ & 887 \\
4 & $+0.12$ & $0.96^{+0.07}_{-0.05}$ & $0.25^{+0.04}_{-0.03}$ & 715 \\
5 & $+0.41$ & $1.15^{+0.23}_{-0.14}$ & $0.28^{+0.06}_{-0.05}$ & 584 \\}
\newcommand{\deltacloud}{0.15^{+0.08}_{-0.05}}

\newcommand{\dslopecloud}{-0.04^{+0.04}_{-0.02}}

\newcommand{\deltalsixeightyeight}{0.40^{+0.15}_{-0.11}}

\newcommand{\splittable}{Ophiuchus & \num{43009} & 0.83 & 0.98 & $0.15^{+0.08}_{-0.05}$ \\
L1688 & \num{6663} & 0.68 & 1.08 & $0.40^{+0.15}_{-0.11}$ \\
L1709 & \num{6721} & 0.64 & 0.76 & $0.12^{+0.16}_{-0.07}$ \\
L1689 & \num{6519} & 0.93 & 0.86 & $-0.07^{+0.18}_{-0.17}$ \\
B44 & \num{11542} & 0.87 & 0.99 & $0.12^{+0.07}_{-0.07}$ \\
North & \num{8908} & 0.50 & 0.63 & $0.13^{+0.10}_{-0.04}$ \\
diffuse tercile 1 & \num{14337} & 0.49 & 0.67 & $0.18^{+0.05}_{-0.14}$ \\
diffuse tercile 2 & \num{14337} & 0.88 & 0.92 & $0.04^{+0.27}_{-0.07}$ \\
diffuse tercile 3 & \num{14337} & 0.87 & 1.04 & $0.17^{+0.07}_{-0.06}$ \\}

\newcommand{\loolow}{0.73}
\newcommand{\loohigh}{1.47}
\newcommand{\loomin}{0.61}

\newcommand{\vthrlow}{0.60}
\newcommand{\vthrhigh}{0.88}
\newcommand{\vslopex}{0.58}
\newcommand{\vslopelo}{0.47}
\newcommand{\vslopehi}{0.71}
\newcommand{\vdeltacloud}{0.16^{+0.04}_{-0.03}}

\newcommand{\deltatemp}{0.35^{+0.07}_{-0.06}}

\newcommand{\vfreedelta}{0.05^{+0.07}_{-0.08}}
\newcommand{\pvfreedelta}{75}

\renewcommand{\dbltopfraction}{0.95}\renewcommand{\dblfloatpagefraction}{0.6}

\begin{document}

\title{Water-ice formation in Ophiuchus follows the local irradiation}

\titlerunning{Water-ice formation in Ophiuchus}
\authorrunning{S. Meingast and Z. Cao}

\author{Stefan Meingast\inst{1}
  \and Zhetai Cao\inst{2}}

\institute{University of Vienna, Department of Astrophysics, T\"urkenschanzstra\ss e 17, 1180 Wien, Austria\\
  \email{stefan.meingast@univie.ac.at}
  \and School of Physics and Astronomy, Beijing Normal University, Beijing 100875, China}

\date{Received 1 October 2026 / Accepted 10 November 2026}

\abstract{
Water ice only  appears on interstellar dust grains beyond a certain depth into a molecular cloud. This depth has so far been quoted as one extinction threshold per cloud. We measured the threshold along \nstars{} sightlines through the Ophiuchus cloud with the ice color-excess method, which converts archival WISE and \textit{Spitzer} photometry of background stars into the peak optical depth of the \SI{3}{\um} water-ice band. We sorted the sightlines by the \SI{12}{\um} surface brightness of the cloud at fixed extinction, a proxy for the ultraviolet field that excites the polycyclic aromatic hydrocarbons at the cloud surface. The onset of ice formation lies at higher extinction toward sightlines with brighter \SI{12}{\um} emission, ranging from $\akz = \SIrange{\thrlow}{\thrhigh}{\mag}$. It increases by \SIrange{\vslopex}{\slopex}{\mag} in \ak per factor of ten in \SI{12}{\um} brightness, depending on the adopted reddening coefficient. Because the \SI{12}{\um}-bright and -faint sightlines lie in different parts of the cloud, we repeated the comparison within each of the five subregions, and four of the subregions show the same trend. The trend also persists among sightlines with a similar amount of diffuse material around them. A later onset in more strongly irradiated gas is expected from photodesorption by external ultraviolet light, although a comparison of its size with models requires a calibration of the \SI{12}{\um} brightness against the strength of the field. The extinction at which water ice appears thus varies with the local environment, which complicates comparisons between clouds. Photometric ice maps resolve this variation across entire clouds, and the SPHEREx and JWST ice surveys can test it spectroscopically.}


\maketitle
\nolinenumbers

\section{Introduction}
\label{sec:introduction}

Water ice is the dominant solid molecular reservoir in molecular clouds and the starting point of the chemistry that ends in comets and planets \citep{Boogert2015}. Its presence is measured through the broad O--H stretching band at \SI{3}{\um} in the spectra of stars behind the cloud. The first systematic surveys already showed that the band only appears above a certain amount of dust extinction, the so-called ice formation threshold. In Taurus, ice absorption starts at $A_V \approx \SI{3.2}{\mag}$ and then grows linearly with extinction \citep{Whittet1988, Whittet2001}. The threshold is understood as the depth at which the far-ultraviolet (FUV) photons that desorb water molecules from the grain surfaces are attenuated enough for freeze-out to prevail \citep{Hollenbach2009, Oberg2009, Cuppen2007}. It marks where the ice mantle becomes thick enough to detect, and grains below it can already carry some ice. Stars and planets form much deeper in the cloud, but ice formation in the irradiated outer layers matters in its own right. Ice-coated grains coagulate more readily than bare ones \citep{Ormel2009}, so grain growth can speed up once ice forms and change the infrared extinction law \citep{McClure2009}. FUV irradiation also processes the ice and creates radicals \citep{Oberg2016}.

Thresholds have been measured in several clouds, and they differ. Lupus and Perseus show ice at $A_V \approx \SIrange{2}{3}{\mag}$ \citep{Boogert2013, Madden2022}, IC~5146 at \SIrange{3}{4}{\mag} \citep{Chiar2011}, and Ophiuchus, exposed to the B stars of Upper Scorpius, at $A_V \approx \SIrange{10}{15}{\mag}$ \citep{Tanaka1990}, with a similarly high threshold for solid CO \citep{Kerr1993}, both measured mostly toward embedded young stars. The differences were attributed to the radiation field of nearby hot stars \citep[e.g.,][]{Tanaka1990, Boogert2013, Madden2022}, and a large scatter of the threshold within Taurus was noted early on \citep{Murakawa2000}. However, each cloud provides essentially a single number, measured with different instruments and extinction laws on a few dozen sightlines at most. Whether the threshold tracks the local radiation field, and by how much, is therefore only inferred from cloud-to-cloud comparisons.

The \ice method \citep{Meingast2025} replaces spectroscopy with broadband photometry. It uses the color $W_1 - I_1$ between the Wide-field Infrared Survey Explorer (WISE) \SI{3.4}{\um} band, which covers the peak of the ice feature, and the \textit{Spitzer} \SI{3.6}{\um} band, which misses most of it. Calibrated against spectroscopic optical depths, this color gives \taumax for every star with archival photometry. With tens of thousands of sightlines through one cloud, the threshold can be measured as a function of the local environment, on one extinction scale and one calibration. In this Letter, we measure the threshold across Ophiuchus as a function of the \SI{12}{\um} surface brightness of the cloud, a proxy for the local ultraviolet field. Throughout, we give extinction in the $K_S$ band and use $A_V \approx 9\,\ak$ \citep{Rieke1985} for literature values.

\section{Data and method}
\label{sec:data}

\subsection{Ice color excess and extinction}
\label{ssec:ice}

We cross-matched the \textit{Spitzer} Infrared Array Camera (IRAC) photometry in $I_1$ of the \textit{Spitzer} Extended Solar Neighborhood Archive \citep[SESNA;][]{Pokhrel2020} for Ophiuchus with unWISE photometry in $W_1$ \citep{Schlafly2019}, the Two Micron All Sky Survey \citep[2MASS;][]{Skrutskie2006}, and Gaia Data Release 3 \citep[DR3;][]{GaiaDR3}, keeping the best counterpart within \ang{;;1}. The ice color excess of a star is $\lam = (W_1 - I_1) - (W_1 - I_1)_0 - \psi\,\ak$, where $(W_1 - I_1)_0$ is the intrinsic color, and $\psi$ is the reddening of the color per unit extinction (Sect.~\ref{ssec:fits}). We converted it into the peak optical depth of the \SI{3}{\um} band with the empirical calibration of \citet{Meingast2025}, $\taumax = -3.51 \ln(1 - 0.92\,\Lambda)$. The extinction toward each star comes from the Herschel and Planck dust opacity map (HP2) of Ophiuchus at \ang{;;36} resolution \citep{Alves2025}, so that \ak measures the full cloud column in the beam around the star. We converted the \SI{850}{\um} optical depth as $\ak = \gamma\,\tau_{850} + \delta$ with $\gamma = \SI{4900}{\mag}$ and $\delta = \SI{-0.07}{\mag}$, the coefficients of the fit against the 2MASS extinction with the method of \citet{Lombardi2014} that is shown in Appendix~A of \citet{Alves2025}; the values are from J.~Alves (priv.~comm.).

We kept stars with clean unWISE flags, photometric errors below \SI{0.1}{\mag} in both bands, no Gaia parallax larger than \SI{7}{mas}, and $I_1 < \SI{15}{\mag}$. The brightness cut removes a color bias of faint stars in bright nebulosity (Appendix~\ref{app:checks}). We also removed young stellar objects (within \ang{;;2} of one listed by \citealt{Dunham2015} or classified as such in SESNA) and duplicated unWISE identifiers. The final sample holds \nstars{} stars behind a field of \ang{5} $\times$ \ang{6}. Appendix~\ref{app:icemap} shows the ice map built from all stars.

\begin{figure}[t]
    \centering
    \resizebox{\hsize}{!}{\includegraphics{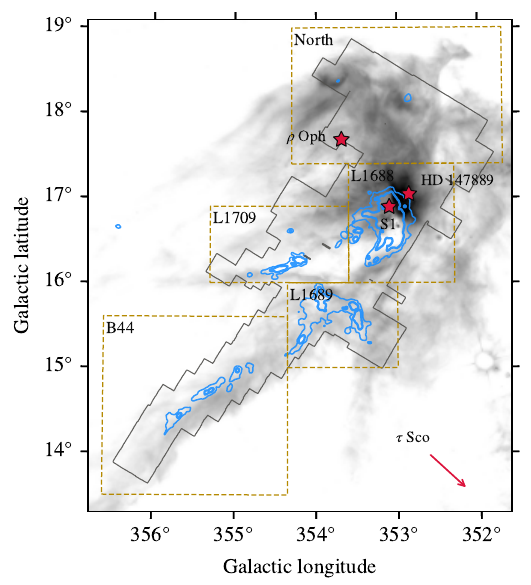}}
    \caption[]{Ophiuchus cloud at \SI{12}{\um} (WSSA, grayscale, arcsinh stretch). The contours show the HP2 extinction map at $\ak = \SIlist{1;2}{\mag}$ (blue), the gray line shows the \textit{Spitzer} coverage, and the dashed boxes outline the five subregions used in the text (B44 is the streamer L1712--L1729). The red stars mark the nearby early-B stars: HD~147889 (B2V) and S1 (B3) in L1688, $\rho$~Oph (B2IV) to the north, and $\tau$~Sco (B0V) \ang{4} south of the field (arrow). $\sigma$~Sco (B1III) lies just beyond the western edge.}
    \label{fig:map}
\end{figure}

\subsection{The environment of a sightline}
\label{ssec:environment}

As a proxy for the FUV field, we used the diffuse \SI{12}{\um} surface brightness $S_{12}$ of the cloud at the position of each star (in \si{\mega\jansky\per\steradian}, smoothed to \ang{;;32} resolution) from the point-source-subtracted WISE Sky Survey Atlas \citep[WSSA;][]{Meisner2014}. Figure~\ref{fig:map} shows this map for our field, together with the five subregions of the cloud that we analyze separately below. The emission in this band comes mostly from polycyclic aromatic hydrocarbons (PAHs) and very small grains, which are heated by single ultraviolet and optical photons \citep{Tielens2008, Draine2007}. For the fields expected in Ophiuchus, their emission is proportional to the intensity of the starlight \citep{Draine2007}, but it also scales with the column of PAHs, and hence with \ak, and it depends on their abundance and charge. To remove the column, we defined the environment of a star as the residual of its \SI{12}{\um} brightness at fixed extinction,
\begin{equation}
    \xii = \log_{10}(S_{12}) - \langle \log_{10}(S_{12}) \rangle_{\ak},
\end{equation}
where the second term is the median of $\log_{10}(S_{12})$ of the stars at the same extinction (a running median over the whole sample, in windows of 2000 stars ordered by \ak), and the residual is in dex. A positive \xii marks a sightline through material that is brighter at \SI{12}{\um} than the cloud average, that is, more strongly irradiated, or richer in small grains. Between the first and 99th percentile of the stars, the residual spans \SI{\xrange}{\dex} (a factor of 13). Within each subregion, the residual varies on arcminute scales.

\subsection{Threshold fits}
\label{ssec:fits}

We describe the ice absorption of a sample of stars by a broken line in \taumax against extinction, $\taumax = s \cdot \max(\ak - \akz, 0)$. Its free parameters are the threshold \akz, the extinction at which detectable ice absorption sets in (hereafter the onset of ice formation), and the growth rate $s$, the optical depth gained per magnitude of extinction above the threshold. We fit the model in color space, so that the reddening term is part of the model and the nonlinear conversion to \taumax did not bias the average \citep[see][]{Meingast2025}. The scatter around the model was described by a Student-$t$ distribution with a free intrinsic width and a free scaling of the catalog errors, which makes the fit robust against variables, blends, and remaining young stars. We measured the reddening coefficient $\psi = \rak$ on the ice-free stars of the full sample ($\ak < \SI{0.5}{\mag}$) and held it fixed, because a threshold, a growth rate, and a reddening slope are degenerate in a single sample. All comparisons below are therefore differential and share the same $\psi$. Uncertainties come from a spatial block bootstrap that resamples HEALPix pixels of \ang{;3.4;} ($N_\mathrm{side} = 1024$) instead of individual stars, so that neighboring stars that share the same map values are drawn together. The same draws were used for all compared samples (a paired bootstrap), so that differences between samples carry their own errors.

\section{Results}
\label{sec:results}

\begin{figure}[t]
    \centering
    \resizebox{\hsize}{!}{\includegraphics{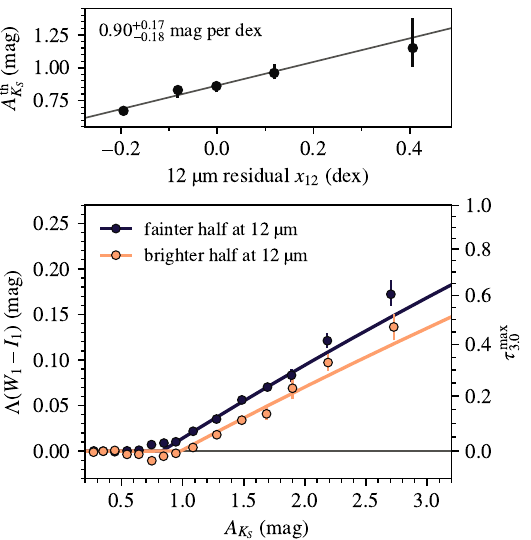}}
    \caption[]{Bottom: Ice color excess against extinction for the \SI{12}{\um}-faint (dark) and the \SI{12}{\um}-bright (orange) half of the sample at fixed extinction. The points are the error-weighted mean of \lam in bins of \ak after 3$\sigma$ clipping, relative to the ice-free stars ($\ak < \SI{0.5}{\mag}$) of the same half, with \SI{68}{\percent} intervals from the spatial block bootstrap. The lines show the broken-line fits of Sect.~\ref{ssec:fits}. The right axis converts the color excess into the peak optical depth of the \SI{3}{\um} band. Top: Threshold of each of the five equal-size bins of the \SI{12}{\um} residual against the median residual \xii of its stars, with the weighted linear relation.}
    \label{fig:main}
\end{figure}

Figure~\ref{fig:main} shows the ice color excess against extinction for the \SI{12}{\um}-bright and the \SI{12}{\um}-faint half of the sample at fixed extinction, each relative to its own ice-free stars. Below $\ak \approx \SI{0.6}{\mag}$, neither half shows ice. Above the onset, the bright half holds less ice at every extinction. Its median color excess is lower by about \SI{15}{\mmag} (\num{0.05} in \taumax) at $\ak = \SIrange{0.8}{1.5}{\mag}$, independent of the broken-line model (Fig.~\ref{fig:checks}) and with about \SI{3}{\mmag} from the nebular color bias of Appendix~\ref{app:checks}, and its threshold is higher by $\Delta\akz = \deltacloud\,\si{\mag}$, positive in all \nboot{} bootstrap replicates, and robust against the choice of $\psi$ (Appendix~\ref{app:fits}).

To measure how the threshold depends on the brightness, we sorted all stars into five equal-size bins by their \SI{12}{\um} residual and fit each bin separately (Table~\ref{tab:bins}). The thresholds rise in order from the faintest fifth of the cloud at \SI{12}{\um} (median $\xii = \SI{\xbinlow}{\dex}$) to the brightest fifth ($\xii = \SI{\xbinhigh}{\dex}$), $\akz = \SIlist{\thrlow;0.83;0.86;0.96;\thrhigh}{\mag}$, with the largest steps at the two ends. The brightest bin holds the weakest signal and has the least-constrained threshold. A weighted line through the five thresholds (top panel of Fig.~\ref{fig:main}) has a slope of \SI{\slopex}{\mag} of \ak per dex of \SI{12}{\um} residual (\SI{68}{\percent} interval \SIrange{\slopelo}{\slopehi}{\mag}). The growth rate, which ranges from $s = \SIrange{\slopeslow}{\slopeshigh}{\per\mag}$, differs between some bins by more than its errors, but not systematically with \xii. Above $\ak \approx \SI{2}{\mag}$, the two halves in Fig.~\ref{fig:main} rise above their broken lines. This is a tentative hint of a second, steeper segment that we do not quantify here. To first order, the \SI{12}{\um} brightness therefore changes only the onset, although the growth rate and the threshold of a bin are correlated in the fit.

As Fig.~\ref{fig:map} shows, the \SI{12}{\um}-bright sightlines cluster in L1688 and in the northern streamers, while the faint ones dominate B44 and L1709. The cloud-wide relation could therefore reflect other differences between these regions instead of the \SI{12}{\um} brightness. To test this, we repeated the two-way split within each subregion, with the residual redefined relative to the stars of that region. Table~\ref{tab:robust} lists the thresholds of both halves in each subregion. The bright half has the later onset in four of the five regions, most clearly in L1688 ($\Delta\akz = \deltalsixeightyeight\,\si{\mag}$) and at lower significance in L1709, B44, and the northern streamers ($\Delta\akz = \SIrange{0.12}{0.13}{\mag}$). In L1689 the two halves have the same threshold within an uncertainty that is too large to tell the region apart from the others.

The relation between the \SI{12}{\um} brightness and the threshold holds within the regions and not between them. The mean thresholds of the five regions do not follow their mean \SI{12}{\um} brightness: the northern streamers are as bright as L1688, but have the lowest threshold of all. The later onset behind bright material appears instead when bright and faint sightlines of the same region are compared (Table~\ref{tab:robust}), that is, on arcminute scales. The slope of the five-bin relation also stays positive when any one region is left out, although for a smaller $\psi$ it depends on L1688 (Appendix~\ref{app:fits}). A split by the dust temperature residual at fixed extinction, the more familiar environment proxy, gives the same picture (Appendix~\ref{app:checks}, Fig.~\ref{fig:temperature}).

The later onset behind bright material might also reflect the geometry of the sightline and not the conditions in which the ice forms. A bright sightline may hold more diffuse material, in an extended envelope or in one of the foreground layers known toward Ophiuchus \citep{Boogert2002}. It then reaches the same total extinction with less column in any single structure, and so with less material deep enough for ice. Its unshielded diffuse gas is bright at \SI{12}{\um}. To test this, we estimated the diffuse column as the 10th percentile of the map extinction within \ang{;10;} of each star and repeated the bright and faint split within terciles of it (Appendix~\ref{app:checks}). The bright half has the later onset in every tercile. The difference is significant only in the tercile with most of the high-extinction stars, where it is as large as in the full sample. The contrast thus survives this test, although the proxy cannot exclude unresolved foreground layers. Where the ice-free material is the irradiated surface of the cloud itself, it is part of the effect, not an alternative to it.

\section{Discussion}
\label{sec:discussion}

The results of Sect.~\ref{sec:results} show that the ice threshold of Ophiuchus is not a single number, but depends on the sightline. The onset of ice absorption comes later behind \SI{12}{\um}-bright material, in the cloud as a whole and within four of its five regions, on arcminute scales, and it persists without the broken-line model and within terciles of the diffuse column. At fixed extinction, the threshold rises with the local \SI{12}{\um} brightness by \SIrange{\vslopex}{\slopex}{\mag} of \ak per dex, from \SIrange{\thrlow}{\thrhigh}{\mag} across the cloud, a factor of 1.7, or $A_V \approx \SIrange{6}{10}{\mag}$. The spectroscopic threshold for Ophiuchus, $A_V \approx \SIrange{10}{15}{\mag}$ \citep{Tanaka1990}, overlaps the upper end of this range and extends above it, while Taurus and Lupus lie far below. The threshold thus varies within one cloud with the same quantity that has been invoked to explain the differences between clouds, the local radiation field. At the depths where stars and planets form, the deficit behind bright material (\numrange{0.05}{0.1} in \taumax, Fig.~\ref{fig:main}) is a small part of the ice column.

The simplest interpretation is photodesorption. In the model of \citet{Hollenbach2009}, ice survives where the FUV flux, attenuated as $\exp(-\gamma_{\rm FUV} \cdot A_V)$ with $\gamma_{\rm FUV} \approx 1.8$, no longer removes water molecules as fast as oxygen atoms arrive. This places the threshold at $A_V^{\mathrm{th}} \simeq 0.56 \ln(\gnull / \gnull^{\rm crit})\,\si{\mag}$, where \gnull is the FUV field in Habing units, and $\gnull^{\rm crit}$ scales with the gas density and the inverse of the desorption yield. At fixed density, the threshold depth grows by $\ln(10) / \gamma_{\rm FUV} = \SI{1.3}{\mag}$ of $A_V$ (\SI{0.14}{\mag} of \ak) per dex of \gnull. The observed rise of the threshold has the predicted sign and, to first order, the predicted form, since only the onset changes. If one dex in \xii corresponds to one dex in the incident field, an assumption we cannot test here, the observed rise of \SIrange{\vslopex}{\slopex}{\mag} of \ak per dex is four to six times the predicted \SI{0.14}{\mag} at fixed density. Scatter in \xii that is unrelated to the field, from the abundance and charge of the PAHs, flattens the measured relation rather than steepening it.

Three effects, all plausible in Ophiuchus, could make the threshold rise by more than \SI{0.14}{\mag} per dex of the field. First, the extinction of a background star measures the total column through the cloud, whereas $A_V^{\mathrm{th}}$ of \citet{Hollenbach2009} is the extinction from the illuminated surface to the point where ice survives. For a cloud lit from both sides, ice can only exist beyond the threshold depth from either face, which doubles the rise of the threshold \citep{Boogert2015}. If the cloud is illuminated from the side, at an angle to our line of sight, as by the Upper Scorpius stars in the west or by HD~147889 within L1688, only part of the measured column, possibly well below half, lies between the illuminated surface and the ice. Second, the small grains set the attenuation of the FUV field, and the high ratio of total to selective extinction toward $\rho$~Oph, $R_V \approx 4$, indicates grains larger than in the diffuse interstellar medium (ISM; \citealt{Cardelli1989, Whittet2001}). For grains grown to micron sizes, the effective $\gamma_{\rm FUV}$ is several times smaller than for diffuse-ISM dust \citep{Heays2017}, which multiplies the depth per dex accordingly. Third, the density enters as $\gnull / n$, so if the \SI{12}{\um}-bright sightlines also pass through less dense gas, the threshold lies deeper in the cloud. None of these factors is measured here, so together, they can account for the observed slope, but do not yet explain it.

The interpretation relies on the \SI{12}{\um} residual as a proxy for the field. Two observations support this reading. The brightest sightlines at fixed extinction are the illuminated surfaces of the cloud, the photodissociation region of HD~147889, where the field reaches several hundred Habing units \citep{Habart2003}, and the flanks of B44 that face the Upper Scorpius stars (Fig.~\ref{fig:map}). Second, the dust temperature residual, a separate measure of the heating, gives the same later onset (Appendix~\ref{app:checks}).

Explanations that do not involve the radiation field are harder to sustain. The dust-based extinction could be biased by the mixing of temperatures along the sightline, which lowers the column most toward the illuminated, \SI{12}{\um}-bright surfaces, or by a higher submillimeter opacity of coagulated grains, which raises it in the dense \SI{12}{\um}-faint gas. Both work against the observed trend. Nebular emission that the $W_1$ photometry does not fully remove would mimic less ice in bright nebulosity. However, the stars with $I_1 < \SI{15}{\mag}$ that we used show no blueward trend with the \SI{12}{\um} brightness (Appendix~\ref{app:checks}). What remains is that the \SI{12}{\um} residual traces both the field and the abundance of small grains. Gas in which PAHs have coagulated onto large grains or have been destroyed is faint at \SI{12}{\um} at fixed column. Grain growth and ice formation are expected to go together \citep[e.g.,][]{Madden2022}, so part of the relation might reflect the state of the grains instead of the field that illuminates them. In either reading, the extinction at which water ice appears is set by the conditions along the sightline.

The thresholds quoted for individual clouds, from \SI{0.25}{\mag} of \ak in Lupus \citep{Boogert2013} to \SI{1}{\mag} or more in Ophiuchus, therefore compare environments as much as clouds, and a single threshold per cloud is a coarse average of a distribution. JWST spectroscopy resolves the ice bands in detail but covers only small fields \citep{McClure2023, Smith2025}, and the SPHEREx ice surveys \citep{Melnick2026, Hora2026} will add spectra of the bands over the whole sky. Photometric ice maps built from the \textit{Spitzer} and WISE archives, as presented here and extended to the Milky Way by \citet{Cao2026}, already cover entire clouds at the resolution of this Letter and are the most immediate way to measure this distribution.

\section{Summary}
\label{sec:summary}

We have inferred the water-ice formation threshold from \nstars{} sightlines through the Ophiuchus cloud with the photometric \ice method, sorted by the \SI{12}{\um} surface brightness of the cloud at fixed extinction. The onset of ice formation lies at higher extinction toward sightlines with brighter \SI{12}{\um} emission, ranging from $\akz = \SIrange{\thrlow}{\thrhigh}{\mag}$. Analyzed separately, four of the five subregions show the same trend, although at lower significance outside L1688, so it does not arise from differences between parts of the cloud. The extinction at which water ice forms thus emerges as a property of the local environment rather than of a cloud, and photometric ice maps can now resolve it across entire clouds. Applied to the archival infrared surveys of the Milky Way, this turns the onset of ice formation into a probe of the conditions inside molecular clouds.

\begin{acknowledgements}
We thank the anonymous referee for a constructive report that improved the clarity of this Letter. This work has made use of data from the European Space Agency (ESA) mission Gaia (\url{https://www.cosmos.esa.int/gaia}), processed by the Gaia Data Processing and Analysis Consortium (DPAC, \url{https://www.cosmos.esa.int/web/gaia/dpac/consortium}). Funding for the DPAC has been provided by national institutions, in particular the institutions participating in the Gaia Multilateral Agreement. This research has made use of Astropy \citep{astropy2022} and of the SIMBAD and VizieR services at CDS, Strasbourg, France. The project, its ideas, and its concepts are the authors' own; generative AI tools (Claude, Anthropic; Codex, OpenAI) were used, under the authors' oversight, for drafting and revising the text, for developing the analysis and figure code, and for internal reviews.
\end{acknowledgements}

\bibliographystyle{aa}
\bibliography{references}

\begin{appendix}
\nolinenumbers
\renewcommand{\topfraction}{0.9}
\renewcommand{\dbltopfraction}{0.9}
\renewcommand{\bottomfraction}{0.6}
\renewcommand{\textfraction}{0.07}
\renewcommand{\floatpagefraction}{0.8}
\renewcommand{\dblfloatpagefraction}{0.8}
\setcounter{topnumber}{3}
\setcounter{dbltopnumber}{2}
\setcounter{totalnumber}{5}

\section{The ice map}
\label{app:icemap}

\begin{figure}[ht]
    \centering
    \resizebox{\hsize}{!}{\includegraphics{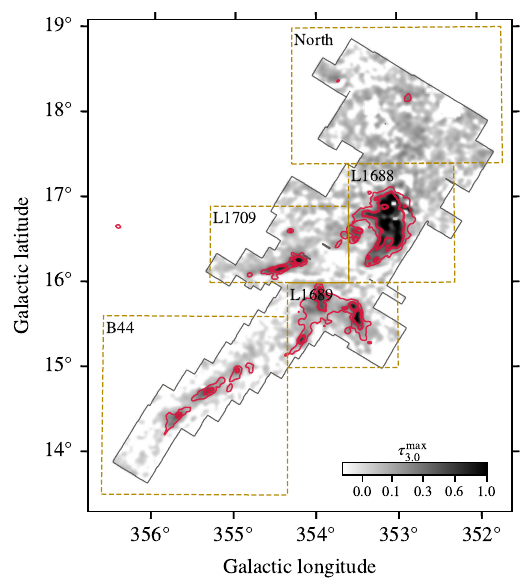}}
    \caption[]{Map of the peak optical depth of the \SI{3}{\um} water-ice band in Ophiuchus from the ice color excess of \num{132483} background stars, smoothed with a \ang{;3;} kernel, within the \textit{Spitzer} coverage (gray outline). Red contours mark $\ak = \SIlist{1;2}{\mag}$ of the HP2 extinction map, and dashed boxes the subregions.}
    \label{fig:icemap}
\end{figure}

Here we provide additional information that is referenced in the main text. Figure~\ref{fig:icemap} shows the ice map of Ophiuchus for orientation. We built the map from all \num{132483} stars of the master catalog that pass the quality cuts of Sect.~\ref{ssec:ice}, except the brightness cut, on a HEALPix grid with \ang{;0.86;} pixels. For each pixel, we averaged the ice color excess of the stars within \ang{;6;} with a Gaussian kernel of \ang{;3;} full width at half maximum and inverse-variance weights from the photometric errors, and then converted the average to \taumax. We applied the conversion to the averaged color excess and not to the individual stars, because \taumax is a nonlinear function of the color excess. The average of noisy individual optical depths would be biased high by about \num{0.01} at the onset of ice formation. The map reproduces the known ice-rich cores of L1688, L1689, and L1709, and it shows the onset of ice along the B44 streamer. We do not use the map for the sightline analysis. All fits are made on the individual stars, and the map only serves to visualize where ice is found.

\section{Fits, reddening coefficient, and robustness}
\label{app:fits}

We fit the broken-line model of Sect.~\ref{ssec:fits} to the observed color $W_1 - I_1$ of each star as $c_0 + \psi \ak + [1 - \exp(-\taumax / 3.51)] / 0.92$ with $\taumax = s \cdot \max(\ak - \akz, 0)$, that is, with the inverse of the calibration of \citet{Meingast2025}. The fit is thus made in the space where the errors are Gaussian, and the intercept $c_0$ absorbs the intrinsic color. The likelihood is a Student-$t$ distribution with four degrees of freedom whose squared scale is $\sigma_{\rm int}^2 + (f \sigma_{\rm phot})^2$, with the intrinsic scatter $\sigma_{\rm int}$ and the error scaling $f$ as free parameters. The catalog errors turn out to be about three times too small at all magnitudes. The likelihood of a broken line is often multimodal in the break. We therefore placed \akz on a grid with a step of \SI{0.01}{\mag}, from \SI{0.1}{\mag} up to the extinction of the 50th most reddened star of the sample, but no higher than \SI{2.5}{\mag}. At least 50 stars thus constrain the growth rate above any trial threshold. For the five bins, the grid ends at \SIrange{2.3}{2.5}{\mag}, well above the fitted thresholds. The northern streamers and the lowest diffuse-column tercile hold only about ten stars above \SI{1}{\mag} per half. Their grids end at \SIrange{0.7}{0.8}{\mag}, which may truncate the upper errors of their thresholds (Table~\ref{tab:robust}). We optimized the remaining parameters for each grid value, and then polished the best solution with all parameters free. Uncertainties come from \nboot{} replicates (100 for the leave-one-region-out and tercile tests) of a spatial block bootstrap with HEALPix blocks of \ang{;3.4;}, which are resampled identically for all samples that are compared. We therefore take differences of thresholds between samples from paired replicates. Tests on synthetic data with asymmetric contaminants recover the threshold to within \SI{0.03}{\mag}.

The reddening coefficient of the color, $\psi = E(W_1 - I_1) / \ak$, cannot be fit together with the threshold and the growth rate in a single sample, because the three parameters are strongly covariant. We therefore measured it on the \nicefree{} stars of the sample with $\ak < \SI{0.5}{\mag}$, where none of the bins shows ice, with the same likelihood and no ice term, and held it fixed at $\psi = \rak$. This value is about twice the \num{0.023} of the extinction curve of \citet{Boogert2011} adopted in the calibration, while the extinction curves used by \citet{Meingast2025} give \numrange{0.02}{0.04} for this color. The excess is not a population effect: the slope is the same in bins of magnitude, of dereddened $J - K_S$ color, and of region, and $J - K_S$ follows the \citet{Boogert2011} curve against the map extinction. The ice-free stars of the two halves give the same $\psi$ to within \num{0.005}. A fit with $\psi$ free in each half trades the threshold against the reddening and reduces the difference between the halves to $\Delta\akz = \vfreedelta\,\si{\mag}$ (positive in \SI{\pvfreedelta}{\percent} of the replicates). This is the degeneracy described above, and it is the reason for holding $\psi$ fixed at a common value.

We can measure $\psi$ only on stars without ice, that is, below the threshold, and we assume the same reddening law at higher column. The result of this Letter does not depend on this assumption, because a different $\psi$ shifts all thresholds together, whereas the difference between sightlines at fixed extinction remains (see below). The absolute value of \taumax does depend on it. Spectra measure the continuum on both sides of the band and so depend much less on this assumption. JWST and SPHEREx spectra of background stars can calibrate the reddening of $W_1 - I_1$ in the ice-bearing gas. A photometric alternative is to measure the law in the near-infrared, where water ice does not absorb, with deep surveys such as VISIONS \citep{Meingast2023}, as done for Orion A by \citet{Meingast2018}, and to extrapolate it into the mid-infrared to obtain the reddening beneath the ice band independently of the ice.

The value of $\psi$ sets the absolute scale of the thresholds. With $\psi = 0.023$, the five thresholds span \SIrange{\vthrlow}{\vthrhigh}{\mag}, and their slope against the \SI{12}{\um} residual drops to \SI{\vslopex}{\mag} per dex (\SIrange{\vslopelo}{\vslopehi}{\mag}). The difference between the halves, $\Delta\akz = \vdeltacloud\,\si{\mag}$, is consistent with the value at $\psi = \rak$. In the leave-one-region-out test, the slope stays positive for every region (\SIrange{\loolow}{\loohigh}{\mag} per dex, \SI{68}{\percent} intervals above \SI{\loomin}{\mag} per dex). With $\psi = 0.023$, it stays positive (\SIrange{0.48}{0.78}{\mag} per dex) without L1709, L1689, B44, or North, but without L1688 the brightest bin is dominated by the northern streamers, whose thresholds are low overall. Its threshold then drops to \SI{0.32}{\mag}, and the weighted slope turns negative.

Table~\ref{tab:bins} lists the fits in the five bins of the \SI{12}{\um} residual, and Table~\ref{tab:robust} collects the two-way splits of the main text. The growth rate above the onset shows no consistent trend with the \SI{12}{\um} residual: the growth rates of the five bins are \SIrange{\slopeslow}{\slopeshigh}{\per\mag} without ordering, and the difference between the halves is $\Delta s = \dslopecloud\,\si{\per\mag}$ cloud-wide and changes sign between regions.

\begin{table}[t]
\caption{Ice thresholds in bins of the \SI{12}{\um} residual.}
\label{tab:bins}
\centering
\small
\renewcommand{\arraystretch}{1.6}
\begin{tabular}{c c c c c}
\hline\hline
Bin & median \xii & \akz & $s$ & $N(\ak > \SI{1}{\mag})$ \\
 & (dex) & (mag) & (\si{\per\mag}) & \\
\hline
\bintable
\hline
\end{tabular}
\tablefoot{Each bin holds one fifth of the sample. \akz and $s$ are the broken-line threshold and growth rate with their \SI{68}{\percent} block-bootstrap intervals; the last column gives the number of stars above $\ak = \SI{1}{\mag}$.}
\end{table}

\begin{table}[t]
\caption{Two-way splits of the sample by the \SI{12}{\um} residual at fixed extinction.}
\label{tab:robust}
\centering
\small
\setlength{\tabcolsep}{4pt}
\renewcommand{\arraystretch}{1.6}
\begin{tabular}{l c c c c}
\hline\hline
Sample & $N$ & \akz faint & \akz bright & $\Delta\akz$ \\
 & & (mag) & (mag) & (mag) \\
\hline
\splittable
\hline
\end{tabular}
\tablefoot{The residual is defined relative to the stars of the same sample. $\Delta\akz$ is the threshold of the \SI{12}{\um}-bright half minus that of the faint half, with the \SI{68}{\percent} interval from the paired block bootstrap. The last three rows split the cloud-wide sample by the diffuse column around each star (the 10th percentile of \ak within \ang{;10;}); only the third tercile holds enough stars above $\ak = \SI{1}{\mag}$ for a constrained fit.}
\end{table}

\section{Checks on the extinction and the photometry}
\label{app:checks}

\begin{figure}[tp]
    \centering
    \resizebox{\hsize}{!}{\includegraphics{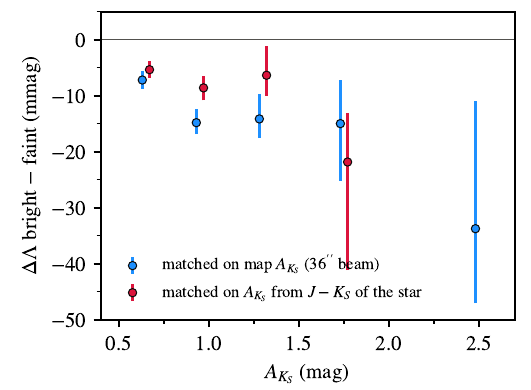}}
    \caption[]{Model-free contrast between the \SI{12}{\um}-bright and the \SI{12}{\um}-faint half of the sample: the median ice color excess of the bright half minus that of the faint half in bins of extinction, each half relative to its own ice-free stars, with \SI{68}{\percent} intervals from the paired block bootstrap. Blue: matched on the beam-averaged extinction of the dust map. Red: matched on the pencil-beam extinction from the $J - K_S$ color of each star.}
    \label{fig:checks}
\end{figure}

\begin{figure}[tp]
    \centering
    \resizebox{\hsize}{!}{\includegraphics{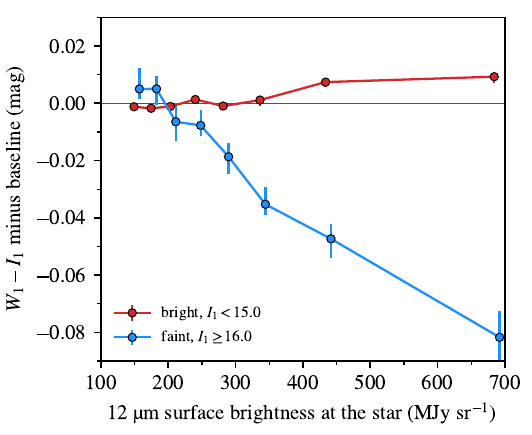}}
    \caption[]{Nebular color bias. Median $W_1 - I_1$ color of ice-free stars ($\SI{0.2}{\mag} \le \ak < \SI{0.5}{\mag}$) relative to the ice-free baseline, against the \SI{12}{\um} surface brightness at the star, for the brightest and the faintest third of the stars in $I_1$.}
    \label{fig:pah}
\end{figure}

\begin{figure}[tp]
    \centering
    \resizebox{\hsize}{!}{\includegraphics{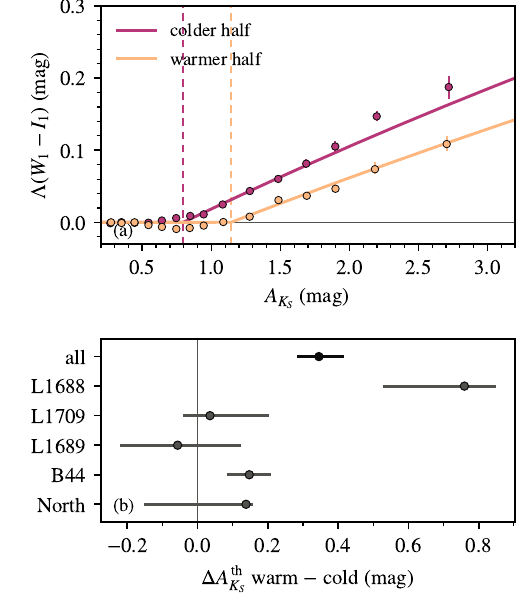}}
    \caption[]{Dust temperature as environment proxy. (a) Ice color excess against extinction for the colder and the warmer half of the sample, split by the dust temperature residual at fixed extinction, as in Fig.~\ref{fig:main}: binned means relative to the ice-free stars of the same half, the broken-line fits, and their thresholds (dashed). (b) Threshold of the warmer half minus that of the colder half for the whole sample and for each subregion, split within the subregion, with \SI{68}{\percent} intervals.}
    \label{fig:temperature}
\end{figure}

This appendix collects the checks referenced in the main text: two on the extinction coordinate, the pencil-beam extinction (Fig.~\ref{fig:checks}) and the diffuse column, one on the photometry, which motivates the brightness cut of Sect.~\ref{ssec:ice} (Fig.~\ref{fig:pah}), and the dust temperature as an alternative environment proxy (Fig.~\ref{fig:temperature}).

\paragraph{Pencil-beam extinction.} The extinction of the dust map is an average over a \ang{;;36} beam, whereas a star samples a pencil beam, the narrow line of sight toward the star itself. Stars that are preferentially detected through holes in a clumpy, illuminated skin would have less extinction than the map says, and hence less ice at a given map extinction. To test this, we computed an extinction for each star from its 2MASS $J - K_S$ color, calibrated against the map on the same stars (a robust line, $J - K_S = 0.72 + 1.31\,\ak$, fit below $\ak = \SI{1.5}{\mag}$). The per-star values scatter around the map by \SI{0.17}{\mag}. At fixed map extinction, the stars behind \SI{12}{\um}-bright material are on average \SI{0.1}{\mag} redder in $J - K_S$ than those behind faint material, so they carry more column than the map assigns to them. Figure~\ref{fig:checks} shows that, matched on the pencil-beam extinction instead of the map, the bright half still holds less ice at every extinction above the onset. The difference is \SIrange{5}{9}{\mmag} up to $\ak = \SI{1.5}{\mag}$, compared with \SIrange{7}{15}{\mmag} on the map extinction, and larger beyond. It is smaller because the scatter of the per-star extinction blurs the onset.

\paragraph{Diffuse column.} Sightlines through \SI{12}{\um}-bright material could carry more diffuse material at the same total extinction, in an extended envelope or in separate foreground layers such as those found toward Elias~29 in L1688 \citep{Boogert2002}. Such a sightline holds less column in any single structure, so its onset appears later even if ice forms at the same depth inside each structure. The test asks whether this diffuse material by itself produces the contrast. It cannot decompose a sightline into physical layers. Diffuse material that is the irradiated surface of the cloud itself is ice-free because of the field and is part of the effect. As a measure of the diffuse column, we took the 10th percentile of the map extinction within \ang{;10;} of each star. This measure correlates with the residual at fixed extinction (Spearman $\rho \approx 0.3$). However, within the tercile of the diffuse column that holds most of the stars above $\ak = \SI{0.8}{\mag}$, the threshold difference between the halves is as large as in the full sample, and the two lower terciles show the same sign (Table~\ref{tab:robust}).

\paragraph{Nebular color bias.} The \SI{3.3}{\um} PAH feature and the emission of hot small grains fall into both bands. With the published response curves, the feature even has slightly more weight in $I_1$ than in $W_1$, so by itself it would make the color redder, not bluer. The two bands differ, however, in resolution. The unWISE point-spread function (about \ang{;;6}) covers more than ten times the solid angle of the IRAC one (\ang{;;1.7}). Nebular structure that the local background estimate does not remove can therefore add more flux to $W_1$ than to $I_1$, most so for faint stars. Stars in bright nebulosity could appear bluer in $W_1 - I_1$, that is, with less ice. Figure~\ref{fig:pah} shows the residual color of ice-free stars against the local \SI{12}{\um} brightness. Stars fainter than $I_1 = \SI{16}{\mag}$ turn bluer by \SI{0.15}{\mag} per dex of surface brightness, whereas stars brighter than $I_1 = \SI{15}{\mag}$ show no blueward trend (\num[retain-explicit-plus]{+0.020 \pm 0.004}\,\si{\mag} per dex). For this reason, the Letter uses only stars with $I_1 < \SI{15}{\mag}$. The halves of Fig.~\ref{fig:main} differ by \SI{0.27}{\dex} in \SI{12}{\um} brightness, so the remaining trend of the bright stars amounts to \SI{5}{\mmag}, with the sign that makes the bright half redder. Since each half is referenced to its own ice-free stars, only the change of this separation with extinction enters the contrast, from \SI{0.30}{\dex} below $\ak = \SI{0.5}{\mag}$ to \SI{0.17}{\dex} at $\ak = \SIrange{0.8}{1.5}{\mag}$: about \SI{3}{\mmag} of the \SI{15}{\mmag} there, or \SI{0.03}{\mag} in $\Delta\akz$. The bright half of Fig.~\ref{fig:main} sits up to \SI{10}{\mmag} below zero at $\ak = \SIrange{0.5}{0.9}{\mag}$, mostly through stars fainter than $I_1 = \SI{13.5}{\mag}$; with $I_1 < \SI{14}{\mag}$ the threshold difference of the halves is \SI{0.11}{\mag} instead of \SI{0.15}{\mag}.

\paragraph{Dust temperature.} The dust temperature of the HP2 map is the more familiar proxy for the environment of a sightline. We split the sample by the temperature residual at fixed extinction in the same way as by the \SI{12}{\um} residual (Sect.~\ref{ssec:environment}). Figure~\ref{fig:temperature} shows that the warmer half has the later onset, $\Delta\akz = \deltatemp\,\si{\mag}$ for the whole sample, clearly in L1688 and B44, with the same sign but within the errors in L1709 and the northern streamers, and with no difference in L1689. The two proxies are strongly correlated (Spearman $\rho = 0.88$). However, the temperature also responds to embedded sources and to the column itself, which is why the main text uses the \SI{12}{\um} emission.
\end{appendix}

\end{document}